\documentclass[aps,twocolumn,prb,preprintnumbers,amsmath,amssymb,superscriptaddress]{revtex4-1}

\usepackage{graphicx}% Include figure files
\usepackage{bm}% bold math
\usepackage{hyperref}
\usepackage{natbib}
\usepackage{float}

\restylefloat{table}

\begin{document}

\title{Measurement of the topological surface state optical conductance in bulk-insulating Sn-doped Bi$_{1.1}$Sb$_{0.9}$Te$_2$S single crystals}

\author{Bing Cheng}
\affiliation{Department of Physics and Astronomy, The Johns Hopkins University, Baltimore, Maryland 21218, USA}

\author{Liang Wu}
\affiliation{Department of Physics and Astronomy, The Johns Hopkins University, Baltimore, Maryland 21218, USA}
\affiliation{Department of Physics, University of California, Berkeley, California 94720, USA}

\author{S. K. Kushwaha}
\affiliation{Department of Chemistry, Princeton University, Princeton, New Jersey 08544, USA}

\author{R. J. Cava}
\affiliation{Department of Chemistry, Princeton University, Princeton, New Jersey 08544, USA}

\author{N. P. Armitage}
\affiliation{Department of Physics and Astronomy, The Johns Hopkins University, Baltimore, Maryland 21218, USA}

\date{\today}

\begin{abstract}
Topological surface states have been extensively observed via optics in thin films of topological insulators. However, in typical thick single crystals of these materials, bulk states are dominant and it is difficult for optics to verify the existence of topological surface states definitively.  In this work, we studied the charge dynamics of the newly formulated  bulk-insulating Sn-doped Bi$_{1.1}$Sb$_{0.9}$Te$_2$S crystal by using time-domain terahertz spectroscopy. This compound shows much better insulating behavior than any other bulk-insulating topological insulators reported previously. The transmission can be enhanced an amount which is 5$\%$ of the zero-field transmission by applying magnetic field to 7 T, an effect which we believe is due to the suppression of topological surface states.  This suppression is essentially independent of the thicknesses of the samples, showing the two-dimensional nature of the transport.  The suppression of surface states in field allows us to use the crystal slab itself as a reference sample to extract the surface conductance, mobility, charge density and scattering rate. Our measurements set the stage for the investigation of phenomena out of the semi-classical regime, such as the topological magneto-electric effect. 
\end{abstract}

\maketitle

Optical measurements have been proposed \cite{Qi08a,Maciejko10a,Tse10a} and shown \cite{Wu16a,Pimenov16a,Okada16a} to be fundamental probes of the non-trivial nature of topological insulators (TI).   To date, however, most such measurements have been on thin films \cite{armitage12,armitage13,armitage15,Post15}.  In contrast, the observation of topological surface states (TSS) in thick TI crystals via optics has been an ongoing challenge.   Some effects such as the half-integer quantum Hall effect may be best demonstrated in thick crystals \cite{Maciejko10a}.  Although there are some optical studies of single crystal TIs, few of them really show explicit signatures of TSS \cite{basov15,basov10,cchomes12,lupi12,Chia13}. This dilemma arises from the principle fact that, although stochiometric TIs are predicted to be semiconductors, their Fermi levels are usually pushed into the bulk bands because of defects introduced in the growth process. A typical example is the Bi$_2$Te$_3$ single crystal where the Fermi level resides in the bulk valence band due to anti-site defects, giving as-grown crystals a large bulk carrier concentration\cite{Chen09,Qu10}. In such TIs, the bulk metallic states are dominant, which interferes with the ability to observe and control the electronic properties of the TSS.  Although surface sensitive probes may give information on the TSS irrespective of the presence of bulk carriers, optical measurements are typically bulk-sensitive and it has been difficult to resolve TSS in the presence of large bulk metallic background signals.

A few bulk-insulating TI crystals have been synthesized recently. For instance, the resistivity of Bi$_{2}$Te$_{2}$Se single crystal exponentially increases with temperature decreasing\cite{ando10}. However, in this compound, the activation energy extracted from resistivity is much smaller than the real band gap. Further optical studies show that impurity bands residing in the band gap contribute most of the low-energy optical responses\cite{cchomes12}.  Moreover, the Dirac point of this compound is below the energy of the maximum of the bulk valence band by 60 meV, making it impossible to study the TSS close to the Dirac point without a contribution from the bulk states.  In a recent advance \cite{cava16}, it has been shown that the material Bi$_{1.1}$Sb$_{0.9}$Te$_2$S doped with a very small percentage of Sn results in almost ideally stable, bulk-insulating, high-crystallinity single crystal.   This material displays a TSS Dirac point, which is energetically isolated from the bulk bands, extremely low bulk carrier densities and transport that is dominated by the surface states below 100 K \cite{cava16}.

In this work, we used time-domain terahertz spectroscopy (TDTS) to study the optical response of this newly discovered bulk-insulating Sn doped TI, Bi$_{1.1}$Sb$_{0.9}$Te$_2$S (BSTS).  The low-energy transmission is above 0.2 at 5 K, which confirms the excellent bulk-insulating properties. With applied magnetic field, the transmission of BSTS shows a small field dependent enhancement.   By virtue of the fact that this enhancement is independent of thickness, we attribute this behavior to the suppression of the TSS.  The suppression of the surface states by field allows us to use the crystal slab itself as a reference to extract the surface conductance directly.   Values extracted for the mobility, scattering rate, and carrier density compare favorably to the thin film counterparts.

\begin{figure}[ht]
\includegraphics[clip,width=3in]{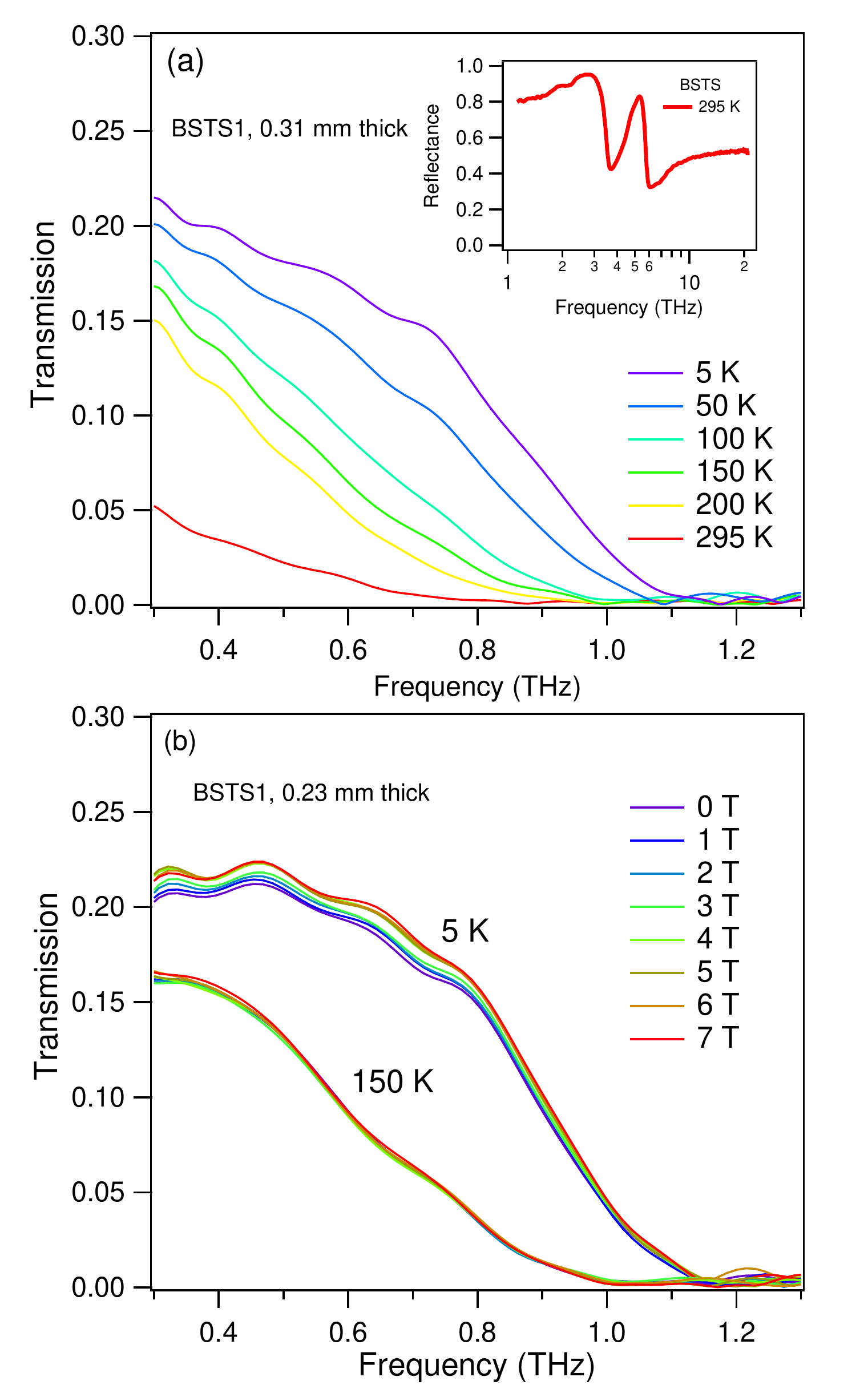}
\caption{(Color online)   (a) Temperature-dependent transmissions of BSTS1. Inset shows far-infrared reflectrace of BSTS at 295 K. (b)  Field-dependent transmission amplitude of BSTS1 at 5 K and 150 K.}
%\end{minipage}
\end{figure}

The Sn-doped single crystals Bi$_{1.08}$Sn$_{0.02}$Sb$_{0.9}$Te$_2$S were grown by the vertical Bridgman technique. Extensive characterization data and design considerations can be found in Ref. \onlinecite{cava16}. BSTS shows a very large insulating-like resistivity\cite{cava16}. At 150 K, its resistivity is approximately 125 $\Omega\cdot$cm. As a point of comparison, the resistivities of other recent ``bulk-insulating" TI single crystals, such as Bi$_{2}$Te$_{2}$Se and Bi$_{2-x}$Sb$_{x}$Te$_{3-x}$Se$_{x}$, are less than 1 $\Omega\cdot$cm at this temperature\cite{ando10,ando11}. Angle-resolved photoemission spectroscopy resolves a clear two-dimensional Dirac surface band and shows that the Fermi level and Dirac point reside inside the band gap. In this work, two pieces of single crystals are studied that we label as BSTS1 and BSTS2.  Considering that the two samples show similar behavior, we mainly focus our attentions on BSTS1.

The original thicknesses of crystals BSTS1 and BSTS2 were 0.60 mm and 0.32 mm, respectively. These crystals were cleaved multiple times with sticky tape to produce different thicknesses of 0.23, 0.31, and 0.60 mm for BSTS1 and 0.26 and 0.32 mm for BSTS2.  Transmission data was collected via TDTS at temperatures down to 5 K. TDTS is a technique in which a femtosecond laser pulse is split along two paths and sequentially excites a pair of photoconductive `Auston'-switch antenna.  A broadband THz pulse is emitted by one antenna, transmits through the reference or sample, and is measured by the other antenna.  Through varying the length difference of the two paths, the electric field of the pulse is measured as a function of time.  In typical TDTS measurements of a homogeneous thin film deposited on a substrate, one ratios the Fourier transform of the pulse transmitted through the sample to the Fourier transform of a pulse transmitted through a nominally identical substrate to get a complex transmission function of the film.  Complex response functions (typically conductivity) can be obtained by analyzing the transmission.  In the present case, THz pulses are transmitted through a thick insulating slab wrapped by two thin conducting surfaces.   Although this is a situation reminiscent of the thin film on a substrate, a difference is that an ideal reference ``substrate" which captures the optical response of the insulating bulk is not readily available. So in this experiment, to get the transmission of the samples, we used vacuum as the reference. This means we need to develop an alternative analysis for the present case to extract the surface response.  This will be discussed in more detail below.

In Fig. 1(a) we show the temperature-dependent transmission amplitude of BSTS1 with a thickness of 0.31 mm. At room temperature, the transmission below 1 THz is very small, but still non-zero. With cooling, the transmission gradually increases and becomes of order 0.2 around 0.3 THz at 5 K. Such a large transmission through a thick sample indicates the insulating nature of the bulk of BSTS. The transmission at 5 K monotonously decreases with increasing frequency. Above 1.2 THz, transmission approaches zero, which means there are strong absorptions above this energy. The inset of Fig. 1(a) shows far-infrared reflectance spectrum of BSTS at room temperature. Two strong phonon peaks are clearly observed in the spectrum at around 1.9 THz and 5.0 THz, respectively. These features, especially the tail of the phonon located at 1.9 THz, result in the strong absorptions observed in the transmission spectra above 1.2 THz.  We believe that the subtle wiggles in the 5 K spectra are artifacts of the transmission measurement because the thick slab sample is not perfectly flat due to the cleaving process. 

In Fig. 1(b) we show the transmission amplitudes at 5 K and 150 K under magnetic fields. At 5 K, with increasing field, the transmission gradually increases. The amount of increase exceeds 5\% of zero-field transmission from 0 T to 7 T at 0.3 THz.  In contrast to 5 K, the increase of transmission at 150 K is much weaker and negligible. We will show below that this increase in transmission at 5 K is independent of the samples' thicknesses and hence we believe it should be almost entirely attributed to the suppression of the TSS conductance.   

To make further progress in the analysis of the TSS properties, we need a model. Unfortunately, directly modeling the transmission with a strict three-layer model is challenging because of the strong absorptions from the tail of the phonon below 2 THz. These strong absorptions dominate the low-frequency response and make it is hard to simulate the insulating bulk due to the exponential sensitivity of the transmission to model parameters.  Instead we employ an alternative method. As mentioned above, in the typical case of TDTS, the transmission of a thin conducting film on a substrate is typically referenced to a nominally identical substrate.   In the present case it makes sense to consider the single crystal slab as a three-layer system which contains a thin metallic surface layer, a thick insulating bulk, and another thin metallic surface layer (see inset to Fig. 3).  Assuming identical properties of the two surface states, this two-side thin film system can be decomposed into two one-side thin film systems. Thus, the transmission of two-side thin film is equal to the transmission of one-side thin film \textit{squared}. With an appropriate reference substrate of index $n$, the appropriate transmission formula is

\begin{equation}
T_{\mathrm{3-layer}} =  \Big( \frac{1+n}{1+n+Z_{0}G}   \Big)^2,
\label{Trans}
\end{equation}

\noindent which is the square of the usual transmission formula \cite{Cheng16}.  Here $n$ is the index of refraction of the reference substrate. Z$_{0}$ is the vacuum impedance, given approximately as 376.7 $\Omega$. This expression assumes identical conductance $G(\omega)$ of the two surfaces.

For an appropriate reference, we may exploit the field dependence of the surface state conductance. In the semiclassical transport regime -- at frequencies less than the transport scattering rate -- one may model the conductance of the surface states as

 \begin{eqnarray}
G_{xx} = \frac{1}{1 + (\mu \textbf{B} )^2}  G(0)  \qquad
\label{Gxx}
G_{xy} = \frac{ \mu \textbf{B} }{1 + ( \mu \textbf{B} )^2}  G(0),
\label{Gxy}
\end{eqnarray}

\noindent where $\mu$ is the mobility of the TSS and $\textbf{B}$ is the applied magnetic field applied perpendicular to the surfaces. $G(0)$ is conductance of a surface state in zero magnetic field.  Note that such expressions ignore all quantum effects of the TSS transport such as weak anti-localization \cite{jun10} or quantized transport \cite{Wu16a} (that could develop at higher fields) as the magneto-conductance effects discussed here are expected to be much larger. The conductances for $R$ and $L$ circularly polarized light -- the polarization eigenstates of a hexagonal crystal under applied field -- are formed by $\hat{G}_{R,L} = G_{xx} \pm iG_{xy}$.  In the high field limit of the semi-classical transport regime, the surface conductance is suppressed because the conductance's spectral weight is moved to higher frequencies and form a cyclotron resonance feature at $\omega_{c} = e  \textbf{B} / m^{*}$. However, for $\omega$ much less that the transport scattering rate $1/\tau$, the frequency independent Eqs. \ref{Gxx} are sufficient.  As the TSS conductance is suppressed at sufficiently high magnetic fields, the transmission at high field $T(B)$ itself may be used as a reference.   The advantage of this trick is that we avoid the problem of simulating the bulk phonon features because they are expected to have negligible field dependence.

\begin{figure}[ht]
\includegraphics[clip,width=3.3in]{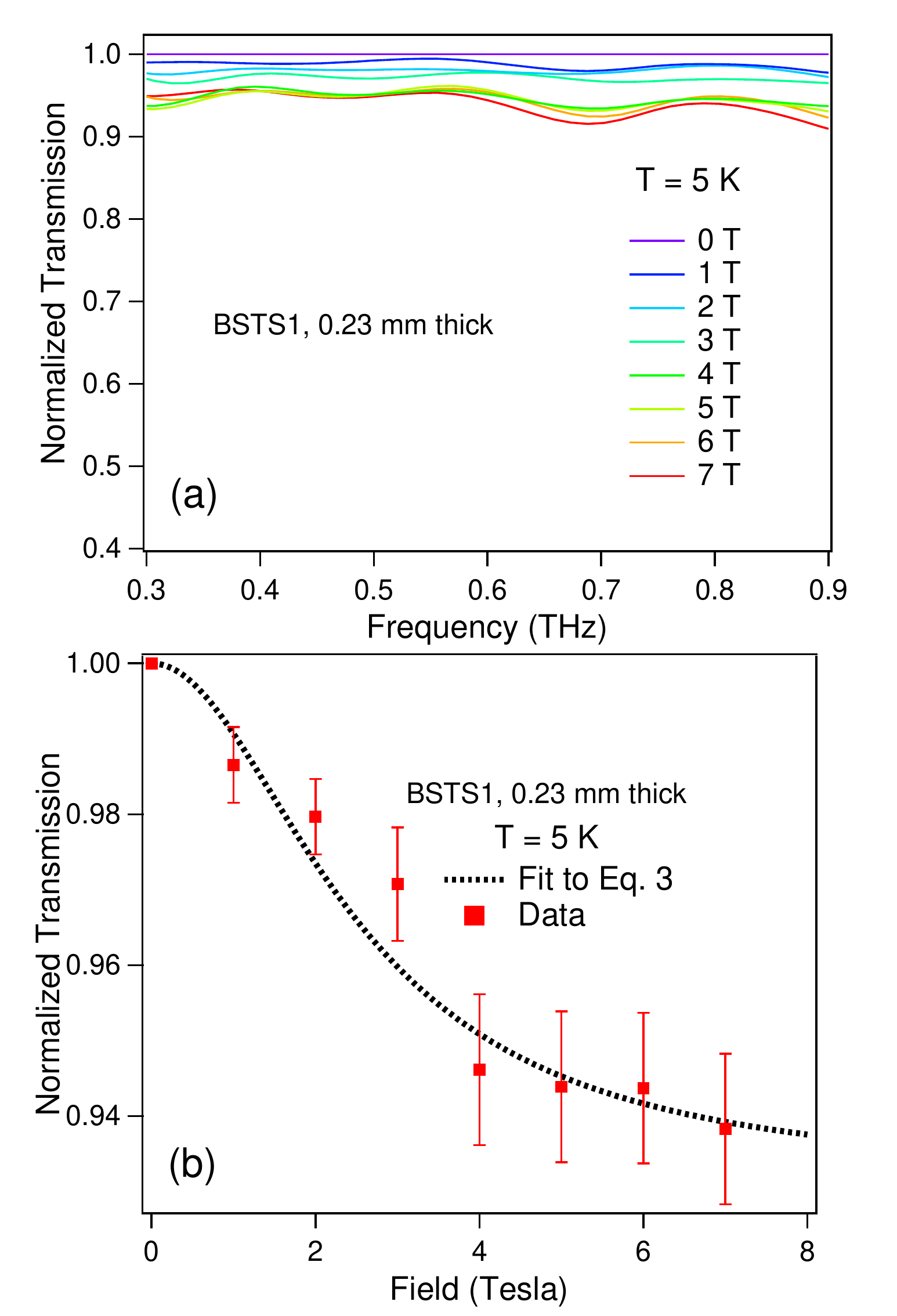}
\caption{(Color online) (a) Transmission at zero magnetic field normalized to the transmission at finite magnetic field $ \frac{T(0)}{T(B)} $   (b) $ \frac{T(0)}{T(B)} $ averaged in a frequency range from 0.30 to 0.90 THz.  The dashed line is a fit to Eq. \ref{NormalizedTransmission}.  }
\label{FigNormTrans}
\end{figure}

\begin{figure*}[t]
\includegraphics[clip,width=7in]{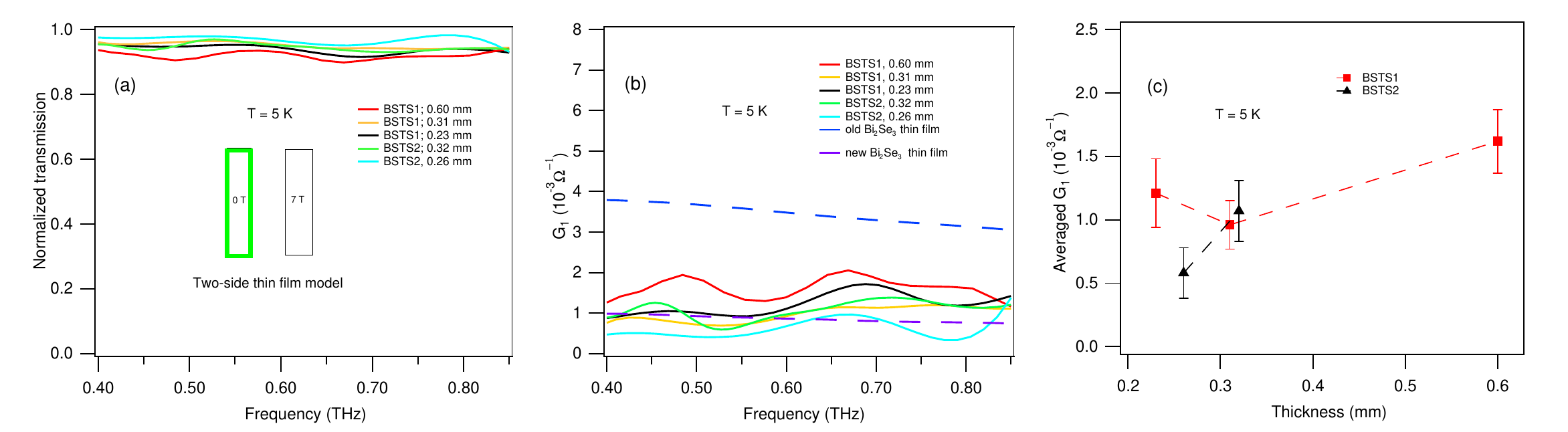}
\caption{(Color online)    (a) Transmissions of 0 T normalized by transmission of 7 T at different thicknesses. Inset:  The green layers represent the thin TI layers of our three layer model.  (b) Real parts of conductance of per TSS extracted from the two-side thin film model. The real part of the conductance per TSS of Bi$_{2}$Se$_{3}$ thin films from Ref. \onlinecite{Wu16a}, \onlinecite{armitage12} are shown for comparison. (c) Thickness-dependent real parts of conductance of per TSS averaged from 0.4 THz to 0.85 THz for BSTS1 and BSTS2. }
\label{FigNormTrans}
%\end{minipage}
\end{figure*}

In Fig. 2(a), we show the results of the ratio of the transmission for the 0.23 mm BSTS1 sample at zero field to the transmission at progressively higher fields.   According to our picture, this quantity decreases at higher fields due to the fact that the TSS conductance is suppressed.   In the high field limit the sample becomes an ideal reference to determine the surface conductance.  However, it may be possible that the surface conductance is not completely suppressed at available laboratory fields.   If the surface conductances are described by Eqs. \ref{Gxx}, the transmission normalized in this fashion $\frac{T(0)}{T(B)}$ is given by,

 \begin{equation}
\frac{T(0)}{T(B)}=  \Big( \frac{2(1 + n + G_R(B))(1 + n + G_L(B))}{(1 + n + G(0)) (2 + 2n + G_{R}(B)+G_{L}(B)) }    \Big)^2.
\label{NormalizedTransmission}
\end{equation}

\noindent In the high field limit of this expression (where $G(B)_{R,L}$ approaches zero), Eq. \ref{NormalizedTransmission} becomes equivalent to Eq. \ref{Trans}.  The free parameters of this expression are the bulk index of refraction $n$ (found to be 13 by independent measurements of the phase delay of the THz pulse across the crystal as compared to a reference), the mobility $\mu$, and the zero field conductance $G(0)$.

As the normalized transmissions shown in Fig. 2(a) do not show any strong frequency dependence (except for the small wiggles that believe are artifacts), we fit the average of normalized transmission in the frequency range from 0.30 - 0.90 THz to Eq. \ref{NormalizedTransmission}. The averaged transmission and fitting are shown in Fig. 2(b).   From the fitting, the mobility is found to be 4000 cm$^{2}$V$^{-1}$s$^{-1}$ and the zero field conductance is approximately 0.0014 $\Omega^{-1}$ for each surface layer.   Similar values were obtained at other thicknesses.

From inspection of Fig. 2(b), the surface-state conductance is found to be largely suppressed by the maximum laboratory field of 7 T. In this regard, we can take transmission at this field as the reference measurement for extraction of the surface conductance and use Eq. \ref{Trans}.   In Fig. 3(a), we plot the zero field transmission which have been normalized to the 7 T data for all five thicknesses.  One can see that the normalized transmission for all samples is essentially identical and thickness-independent, which means that the dissipation is two-dimensional and presumably arises from the TSS.  The similar observation of thickness-independent transmission allowed us to identify TSS transport in the previous work on TI thin films\cite{armitage12}.

Fig. 3(b) shows the real part of the conductance per surface layer at five different thickness extracted from Eq. \ref{Trans}.  Consistent with our assumptions of $\omega \ll 1/\tau$, the surface conductance is indeed flat in our spectral range.  Again, we believe the wiggles, which are now more apparent in Fig. 3(b), are artifacts from the slab geometry and the sample's imperfect flatness. Alongside the extracted surface conductance, we also show optical data from our previous studies on Bi$_2$Se$_3$ thin films. The older generation film had a large charge density with E$_F$ almost 350 meV above the Dirac point which is nearly in the bottom of the conduction band, but still had transport dominated by the TSS\cite{armitage12,oh12}. The second film is grown with a method that results in a true bulk insulator\cite{oh12,oh15}, is more comparable to the present generation of BSTS crystals and has an E$_F$ approximately 50 meV above the Dirac point. The optical conductance of Bi$_2$Se$_3$ thin films show smaller scattering rate than that of BSTS single crystals, which is consistent with the fact that the thin films have nominally perfect stoichiometry. We also plot real parts of conductances averaged from 0.40 THz to 0.85 THz as a function of thickness in Fig. 3(c). The averaged conductances do not scale with thickness and are all close to the value of 0.001 $\Omega^{-1}$which is reported in Ref. \onlinecite{cava16}. These results further support our point that the conductances we extracted come from topological surface states.

Combining the results above, we can estimate the Fermi energy E$_{F}$ and scattering rate $\gamma$ in BSTS. By using Eq. \ref{NormalizedTransmission} to fit field-dependent normalized transmission of BSTS1 with a thickness of 0.23 mm, we obtained zero-field conductance G(0) = 0.0014 $\Omega^{-1}$ and mobility $\mu$ = 4000 cm$^{2}$V$^{-1}$s$^{-1}$. From the expression G(0) = $n_{2d}$e$\mu$, we can derive the charge density of TSS $n_{2d} = 2.2 \times 10^{12} \:cm^{-2}$. The Fermi wave vector $k_{F}$ and $n_{2d}$ are related by the formula $n_{2d}$ = $k_{F}$$^{2}$/4$\pi$, which yields the Fermi wave vector $k_{F}$ = 0.05 \AA$^{-1}$.  With a linear approximation to the massless Dirac dispersion and a Fermi velocity $v_{F}$ of 4 eV$\cdot$ \AA \cite{cava16}, the Fermi energy E$_{F}$ in BSTS1 at 0.23 mm thick is estimated to be 200 meV above the Dirac point.  This extracted Fermi energy is close to, but slightly higher than the 120 meV value for crystals grown by the same method reported in Ref. \onlinecite{cava16}. However, considering the conduction band is found to be 230 meV above the Dirac point, the Fermi level extracted by our study resides inside the band gap, consistent with our transmission measurements. We can estimate the effective mass through the formula m$^{*}$v$_{F}$ = $\hbar$k$_{F}$, which yields an effective transport mass of m$^{*}$ = 0.06 m$_{0}$ (where m$_{0}$ is the mass of electron). Then putting the effective mass m$^{*}$ into the equation $\mu$ = e$\tau$/m$^{*}$, we finally extract out the scattering rate $\gamma$ = 1/2$\pi$$\tau$ = 1.2 THz. This number is larger than our available spectral range justifying our approximation of a frequency independent conductance and demonstrating the self consistency of our analysis. 

In conclusion, we used TDTS to investigate thick bulk-insulating TI Sn doped Bi$_{1.1}$Sb$_{0.9}$Te$_2$S single crystals and verified their excellent bulk insulating features. We studied their optical responses under magnetic field and found a clear signature of TI surface states. To extract the TSS conductance, we develop a three-layer model, in which the transmission through the single crystal in high magnetic fields is used as a reference.  The optical conductance extracted per surface is of order 0.001 $\Omega^{-1}$.  Values extracted for the mobility, scattering rate, and carrier density compare favorably to the thin film counterparts.  Our measurements set the stage for further experiments that may push these materials out of the semi-classical transport regime and investigate the quantum correlations that should manifest in this class of materials.

We would like to thank N. J. Laurita for sharing some of his Igor analysis routines.  THz experiments were supported by the Army Research Office Grant W911NF-15-1-0560.   Work at Princeton was supported by the ARO MURI on topological insulators, grant W911NF-12-1-0461.

%\vspace{-7mm}

\end{document}